\documentclass[11pt]{article}
\usepackage{moriond,epsfig}

\def\be{\begin{equation}}
\def\bq{\begin{equation}}
\def\ee{\end{equation}}
\def\bea{\begin{eqnarray}}
\def\eea{\end{eqnarray}}

\newcommand{\fr}[2]{\frac{#1}{#2}}

\begin{document}
%%%%%%%% HAS BEEN SUPPRESSED \vspace*{4cm}
\title{
ON CLASSICAL FLUCTUATIONS OF BALLISTIC CONDUCTANCE
}

\author{\underline{P.G. SILVESTROV}}

\address{Instituut-Lorentz, Universiteit Leiden, P.O. Box 9506, 2300 RA
Leiden, The Netherlands}

\maketitle\abstracts{
Universal conductance fluctuations~\cite{UCF} in disordered systems are one of the
most known quantum mesoscopic effects. For ballistic cavity with smooth confining 
potential however, one should observe a much larger classical sample-to-sample 
conductance fluctuations.
It is shown, how bending of the phase space in case of chaotic dynamics leads
to 
%nontrivial 
additional enhancement of such fluctuations.
}

\section{Introduction}

Recently a method for investigation of classical to quantum crossover in
ballistic transport was proposed~\cite{Sil03L,Sil03}.
It was expected to become an alternative of averaging over fictitious disorder
used before~\cite{Ale96} in order to mimic the effect of classical chaos.
One advantage of the method was a possibility of explicit accurate construction
of solutions of the Schr\"{o}dinger equation built from
short classical trajectories (shorter than the Ehrenfest time
$\tau_E\propto\ln\hbar$). A disadvantage was conjecturing the
validity of Random Matrix Theory~\cite{Bee97}~(RMT) to describe
the effect of longer trajectories.
Extensive numerical
simulations were performed~\cite{all1,all2} in order to verify the
possibility of using RMT for only a part of phase space. 
Still the understanding of surprising accuracy of effective RMT remains a challenge.
In this note however we make a step back and consider the pure classical
contribution to sample-to-sample conductance fluctuations.
These fluctuations, being $\hbar^{-1}$ times larger~\cite{all2} than the famous 
universal conductance fluctuations~\cite{UCF}, may be seen if by means of external
gates one will change the chaotic dynamics in the cavity while keeping constant
openings to 2DEG.
In all existing analysis the chaos was taken into account via
exponentially fast stretching and squeezing of areas in phase
space. Here we go beyond this linear stretching approximation and show
how anomalously large ``fluctuations of the conductance fluctuations'' 
appear due to bending of phase space.

\section{Open billiard}

Classical trajectories appear in a quantum mechanical scattering problem through
the semiclassical wave function (for two spatial dimensions $x,y$ and 
smooth confining potential $U(x,y)$)
\begin{equation}
\psi(x,y)=\sum_{\sigma}\sqrt{\rho_{\sigma}(x,y)}\exp[i
{\cal S}_{\sigma}(x,y)/\hbar] \ . \label{psidef}
\end{equation}
Here the action ${\cal S}_{\sigma}$ and density
$\rho_{\sigma}$ solve the Hamilton-Jacobi and continuity equations
\begin{equation}
|\nabla {\cal S}|^{2}=2m(E_{F}-U)\;\;,\;\;\nabla\cdot(\rho\nabla
{\cal S})=0.
\label{Hamilton}
\end{equation}
The action is multivalued and the index $\sigma$ labels the different
sheets.
A family of trajectories described by eqs.~(\ref{Hamilton}) forms a tube,
as is shown in the figure~1,~left.
The requirement that $\psi$ is single-valued
imposes a quantization condition,
\begin{equation}
\oint p_{y}dy=(n+1/2)h,\label{quantization}
\end{equation}
where the integral is taken over any contour enclosing the tube.

A complete description of the classical motion in the billiard is given by the
surface of section shown in Fig.~1,~right.
The injected beam crosses the section for the first time over an
area ${O_{\rm initial}}$.
Further crossings consist of increasingly more elongated ($\propto e^{\lambda t}$)
areas ${O}_j$.
The fifth crossing is shown in the figure.
To leave the billiard (through the right contact) without further
crossing of $b$ a particle should
pass through an area $O_{\rm final}$.
Consequently crossings of ${O}_j$ with $O_{\rm final}$ indicate the trajectories
leaving the billiard without further crossings with $b$ and essentially  without
quantum interference. Following ref.~\cite{Sil03} we will call these overlaps of ${O}_j$ and
$O_{\rm final}$ the transmission bands.
Due to~(\ref{quantization})
the conductance is given by the sum of areas
of intersection $G=\sum \oint {\bf p}\cdot d{\bf r}/h$. This counting of areas
allows one to calculate the conductance as long as individual
areas remains larger than~$h$, which is the case for times shorter than $\tau_E$.
On the other hand, sample specific fluctuations of the biggest areas of intersection
will lead to large, classical in essence, fluctuations of the conductance.

\begin{figure}
\psfig{figure=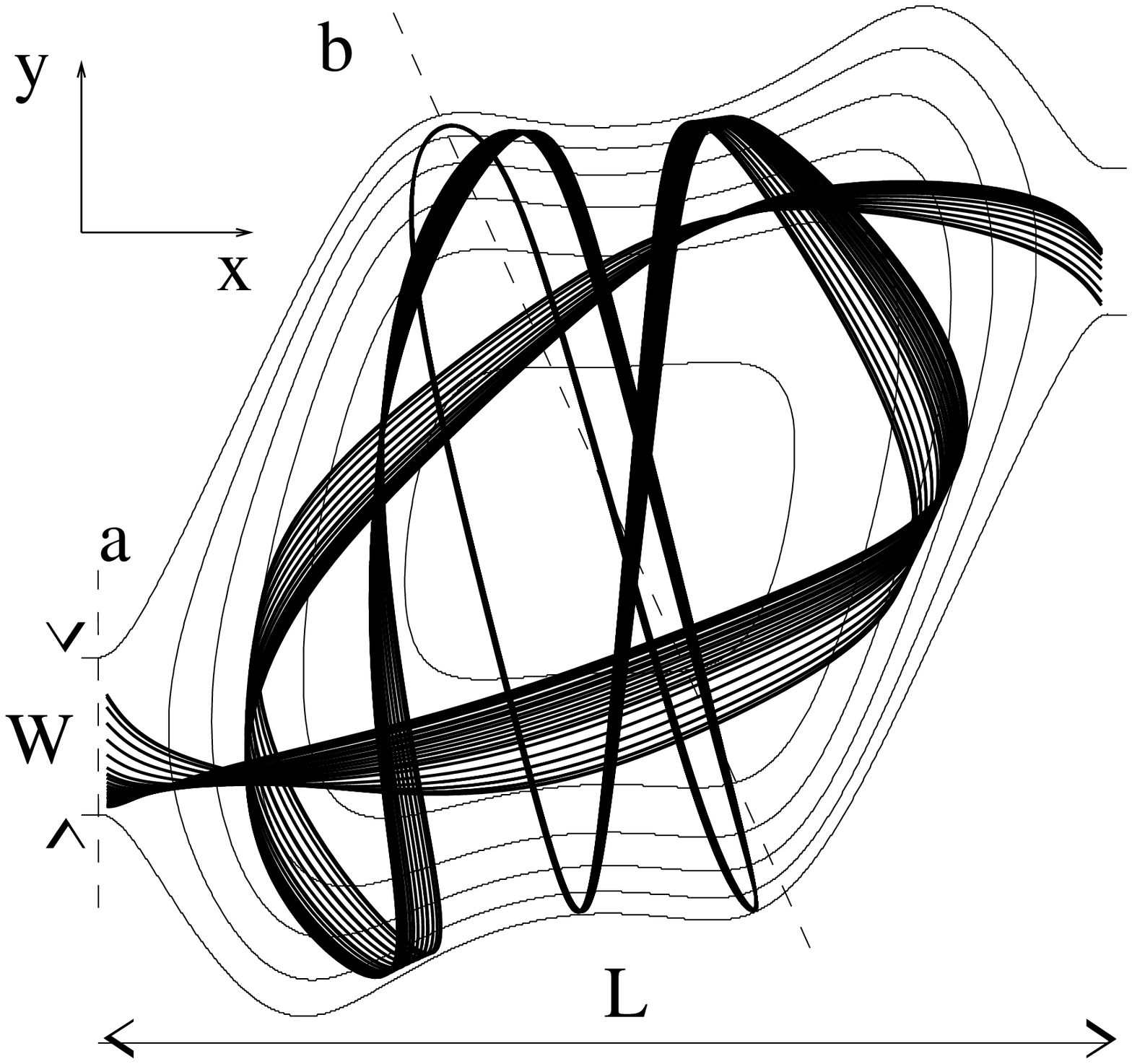,height=2.4in}
\vspace{-2.5in}
\begin{flushright}
\psfig{figure=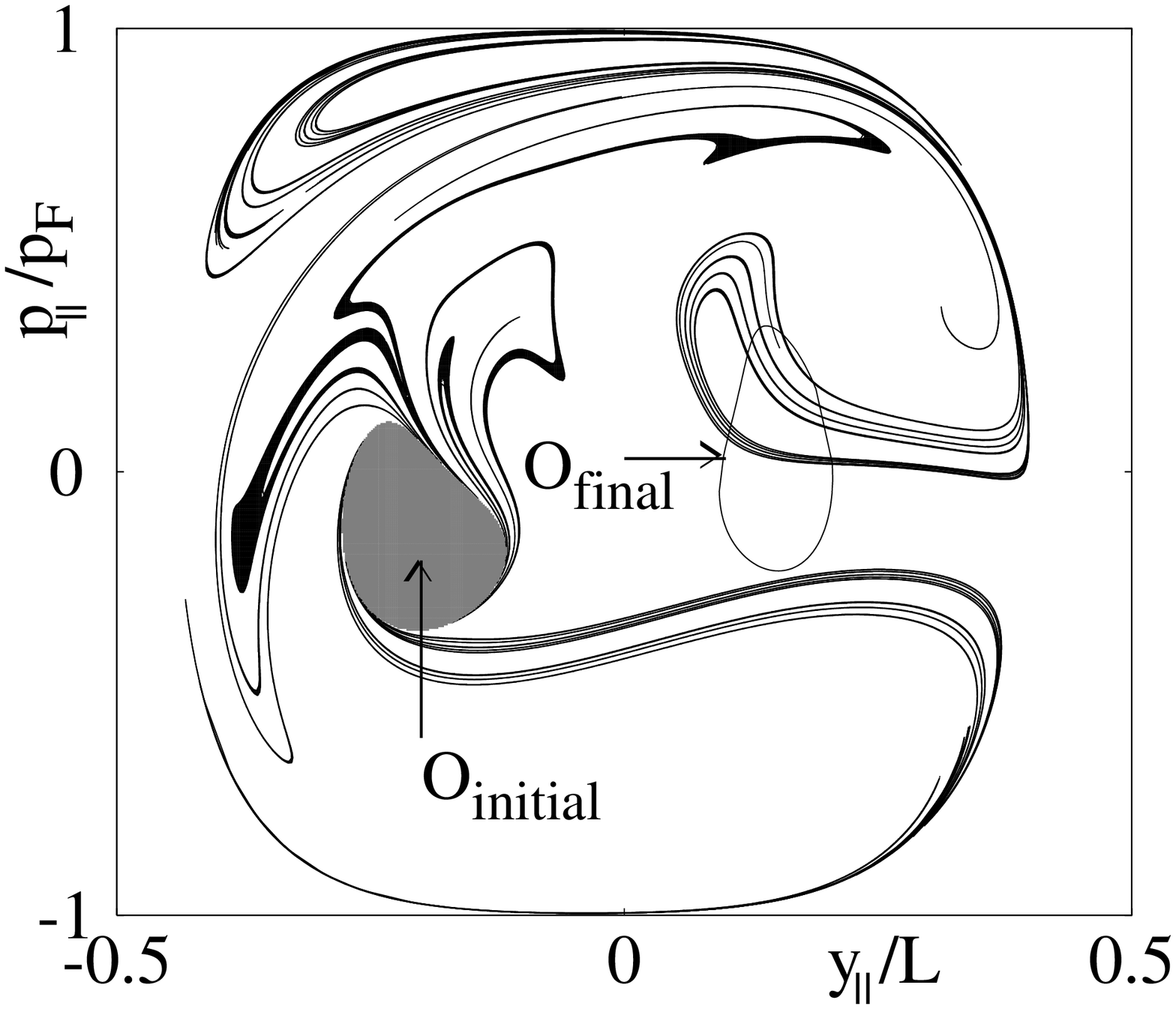,height=2.6in}
\end{flushright}
\caption{Left:~Selected equipotentials of the electron billiard
and a flux tube of transmitted trajectories
representing the spatial
extension of a fully transmitted scattering state.
The outer equipotential defines the area
which is classically accessible at the Fermi
energy.
The motion in the closed billiard is chaotic with Lyapunov
exponent~$\lambda$. Right:~Section of phase space in the middle of the billiard (line $b$ in
left picture).
Axes are the component of coordinate and momentum
along this line.
The area $O_{\rm initial}$ is the position of the first crossing of this surface of
section by {\it all} possible at the Fermi energy injected trajectories.
Elongated black areas $O_j$ show the positions of the 5-th crossing of the
injected beam with the surface of section.
Points inside $O_{\rm final}$ leave the billiard without further crossing
of line $b$.
\label{fig:path}}
\end{figure}

\section{Quantum map model for ballistic cavity}

Since the conductance is determined by the single
surface of section we may investigate
the general features of chaotic billiards by looking at the simple open map
model~\cite{Oss03,all1,all2}.
An example of such map is given by the quantum kicked rotator ($p+2\pi\equiv p, x+2\pi
\equiv x$)
 \bq\label{rotat} H=\fr{\hat{p}^2}{2}
+K \cos(x)\sum_n\delta(t-n) \ , \ \hat{p}=\fr{\hbar}{i} \fr{d}{dx} .
\ee
Choosing a special value of the Plank constant $\hbar\equiv
2\pi/M$ allows to work with a
finite dimensional Hilbert space in both coordinate $x_k=k\hbar$,
and momentum $p_m=m\hbar$ representations,
$k,m=1,2,...M$. This greatly simplifies the numerical calculation of the
Floquet operator
 \bq\label{Fl1}
U=e^{-i\hat{p}^2/2\hbar}e^{-iK\cos(x)/\hbar} .
 \ee
 The classical equations of motion for the kicked rotator are
 \bq \left\{\begin{array}{cc}
p_{n+1}=p_n+K\sin x_n\\
x_{n+1}=x_n+p_{n+1}
\end{array}\right. \ .
 \ee
The $S$-matrix for open kicked rotator is given by~\cite{Oss03,Fyo00}
 \begin{eqnarray}\label{Smat}
S= P(1-UQ)^{-1}UP^T
= PUP^T+PUQUP^T+PUQ UQ
UP^T+... \ .
 \end{eqnarray}
Here $Q=1-P^T P$, and
a rectangular $2N\times M$ matrix $P$ describes coupling to the leads.
Two openings are placed at $x_1<x<x_2$~(injected-reflected) and
$x_3<x<x_4$~(transmitted), where $x_2-x_1=x_4-x_3=w$.
The only nonzero elements of the projection matrix $P$
are $P_{i,i+I_{in}}=P_{i+N,i+I_{out}}\equiv 1$, where $ I_{in}= Mx_1/2\pi$,
$I_{out}= Mx_3/2\pi$ and $0<i\le N$.

The dimensionless conductance
follows from the Landauer formula $G=\mbox{Tr} T$,
where the transmission matrix $T=t^+ t$ is determined by the upper-right $N\times N$
sub-block $t$ of the $S$-matrix.
In our classical limit however calculation of the trace of transmission matrix
reduces to the simple counting of areas of intersections in the $x,p$ plane.

\section{Sample specific conductance fluctuations}

For kicked rotator (\ref{rotat})
the developed chaos corresponds effectively to $K\gg 1$
(the Lyapunov exponent is $\lambda\approx\ln(K/2)$).
Therefore the calculation in this section will be done in the leading order
in $1/K$.
Assuming the ergodic motion we may find the averaged number of transmitted
or reflected channels at the $n$-th iteration of the map
\bq\label{dwell}
\overline{T_n}=\overline{R_n}=\fr{Nw}{2\pi}\exp[-(n-1)w/\pi] \ .
\ee
We are now interested in the fluctuations of these numbers caused by the variations
in the leads position. The fluctuations are naturally more pronounced for small $n$.
In our particular example however the first iteration $n=1$ turns out to be 
trivial. The image of initially injected beam of particles after one iteration 
covers the area between two parallel lines $p+x_1<x<p+x_2$.
The numbers of transmitted and reflected
channels are given by the eq.~(\ref{dwell}) and do not fluctuate.

\begin{figure}
\begin{center}
\psfig{figure=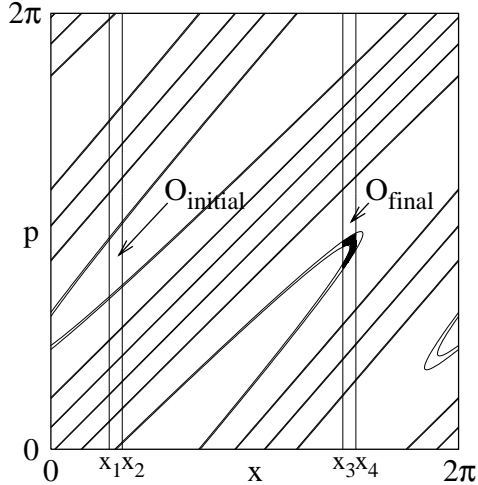,height=2.5in}
\end{center}
\caption{ Second iteration of the map. Areas $O_{\rm initial}$ and 
$O_{\rm final}$ correspond to $x_1<x<x_2$ and $x_3<x<x_4$. Tilted narrow lines
are the second image of the injected area $O_{\rm initial}$. Their overlap with
$O_{\rm initial}$ and $O_{\rm final}$ defines the transmission~(reflection) bands.
The black area shows the biggest transmission band.}
\vspace{-.2cm}
\end{figure}

Sample specific fluctuations of conductance first appear at the second iteration, as
is illustrated by Fig~2.
The typical number of channels in the transmission band is of the order of
$\sim Nw/K$. However, if the lead will be placed close to one of the turning points
of the bent image of the opening, the number may become as large as $N\sqrt{w/K}$.
For narrow openings $w\ll 1$ and $K\gg 1$, these largest transmission bands cover 
an area between two inclined parabolas,
 \begin{eqnarray}\label{second}
p=x\pm \sqrt{2x/K} \ \ \mbox{and} \ \
p=x\pm \sqrt{2(x-w)/K}  \ .
 \end{eqnarray}
A simple integration allows to find the areas of intersection for different relative
positions of the openings.
This results in a probability distribution for large numbers of transmitted channels
\bq\label{distrib}
P(T_2)=\fr{4N^2w^2}{\pi^3KT_2^3} \ , \ \ \mbox{\rm where} \ \
\fr{Nw}{\sqrt{K}}<T_2<\fr{N\sqrt{w}}{K} \ .
\ee
The same probability distribution describes fluctuations of the number of reflected
channels $R_2$.

Contributions to conductance coming from the transmission bands which appear at
the third and higher iterations of the map contain extra factors $K^{-1}$ and 
their fluctuations may be neglected. Thus the dimensionless conductance has a form
$G=T_2+(N-T_2-R_2)/2$ and the
fluctuation of conductance is $\delta G\approx T_2/2-R_2/2$.
The power law tail of distribution (\ref{distrib}) leads to the additional 
$\sim\ln(w)$ enhancement of the second moment
 \bq\label{var}
\langle \delta G^2\rangle =
\fr{w^2N^2}{K\pi^3}\ln\left(\fr{1}{w}\right) .
 \ee
This logarithmic dependence on $w$ ($w\ll 1$)
may be responsible for the deviations from the scaling
$\langle \delta G^2\rangle \sim w^2N^2$ observed in numerical simulations~\cite{all2}.

Numerical coefficients in eqs.~(\ref{distrib},\ref{var}) are specific for the
model~(\ref{rotat},\ref{Smat}). In more realistic case of a ballistic quantum dot
of the size $L$ coupled to the leads by two $N$-mode
contacts eqs.~(\ref{distrib},\ref{var}) transform into
\bq\label{res}
P(\delta G)\sim \fr{N^4}{(k_F L)^2}\fr{1}{\delta G^3} \ \ , \ \
\langle \delta G^2\rangle \sim  \fr{N^4}{(k_F L)^2}\ln\left(
\fr{k_F L}{N}
\right) \ .
\ee
Several experiments~\cite{Obe02,Rit03} were performed recently aimed to observe quantum
to classical crossover in quantum dots. In the experiment of this kind one is able to
vary manually the conductance of the contacts and the shape of the dot.
Further improvement of this technic may lead to experimental verification
of our results.

\section*{Acknowledgments}
Discussions with C.W.J.\ Beenakker and J.\ Tworzyd\l o are greatly appreciated.
This work was supported by the Dutch Science Foundation NWO/FOM.

\end{document}